# Structural Color Production in Melanin-based Disordered Colloidal Nanoparticle Assemblies in Spherical Confinement[†]

Anvay Patil,[1,‡] Christian M. Heil,[2,‡] Bram Vanthournout,[3] Markus Bleuel,[4,5] Saranshu Singla,[1] Ziying Hu,[6] Nathan C. Gianneschi,[6,7] Matthew D. Shawkey,[3] Sunil K. Sinha,[8] Arthi Jayaraman,[2,9,*] and Ali Dhinojwala[1,*]

[1]*School of Polymer Science and Polymer Engineering, The University of Akron, 170 University Ave, Akron, Ohio 44325, USA.*

[2]*Department of Chemical and Biomolecular Engineering, University of Delaware, 150 Academy St, Newark, Delaware 19716, USA.*

[3]*Evolution and Optics of Nanostructures Group, Department of Biology, Ghent University, Ledeganckstraat 35, Ghent 9000, Belgium.*

[4]*NIST Center for Neutron Research, National Institute of Standards and Technology, Gaithersburg, Maryland 20878, USA.*

[5]*Department of Materials Science and Engineering, University of Maryland, 4418 Stadium Dr, College Park, Maryland 20742, USA.*

[6]*Department of Chemistry, Northwestern University, Evanston, Illinois 60208, USA.*

[7]*Department of Materials Science and Engineering, Department of Biomedical Engineering, Department of Pharmacology, International Institute of Nanotechnology, Simpson-Querrey Institute, Chemistry of Life Processes Institute, Lurie Cancer Center, Northwestern University, Evanston, Illinois 60208, USA.*

[8]*Department of Physics, University of California San Diego, 9500 Gilman Dr, La Jolla, California 92093, USA.*

[9]*Department of Materials Science and Engineering, University of Delaware, 201 DuPont Hall, Newark, Delaware 19716, USA.*

[‡]A.P. and C.M.H. contributed equally to this work.

[*]Corresponding authors: ali4@uakron.edu (A.D.), arthij@udel.edu (A.J.)

[†]Electronic supporting information (ESI) available upon request or on the published article's webpage (see DOI)

**Table of Contents**

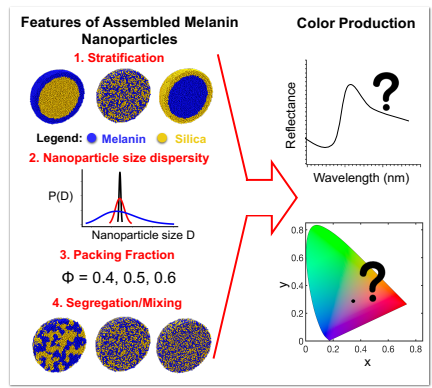

This work combines molecular dynamics and finite-difference time-domain computational approaches to study structural color production from melanin nanoparticle-based supra-assemblies. The fundamental knowledge of how melanin controls and alters color will be valuable for engineering synthetic optical materials for applications in paints, coatings, cosmetics, and food colorings.

**Abstract**

Melanin is a ubiquitous natural pigment that exhibits broadband absorption and high refractive index. Despite its widespread use in structural color production, how the absorbing material, *melanin*, affects the generated color is unknown. Using a combined molecular dynamics and finite-difference time-domain computational approach, this paper investigates structural color generation in one-component melanin nanoparticle-based supra-assemblies (called supraballs) as well as binary mixtures of melanin and silica (non-absorbing) nanoparticle-based supraballs. Experimentally produced one-component melanin and one-component silica supraballs, with thoroughly characterized primary particle characteristics using neutron scattering, produce reflectance profiles similar to the computational analogues, confirming that the computational approach correctly simulates both absorption and multiple scattering from the self-assembled nanoparticles. These combined approaches demonstrate that melanin's broadband absorption increases the primary reflectance peak wavelength, increases saturation, and decreases lightness factor. In addition, the dispersity of nanoparticle size more strongly influences the optical properties of supraballs than packing fraction, as evidenced by production of a larger range of colors when size dispersity is varied versus packing fraction. For binary melanin and silica supraballs, the chemistry-based stratification allows for more diverse color generation and finer saturation tuning than does the degree of mixing/demixing between the two chemistries.

# 1. Introduction

Structural colors are produced by constructive interference of specific wavelengths of light as it moves through a nanostructured material.[1] Typical routes for producing structural colors involve the use of periodic nanostructures (like photonic crystals)[2] or short-range ordered structures (like disordered arrays).[3] Such ordered and disordered morphologies can be readily achieved via assembly of spherical colloidal nanoparticles in confined geometries.[4] The highly-ordered materials tend to produce iridescent structural colors that change depending on the angle of viewing and orientation while disordered or amorphous materials tend to be non-iridescent.[5] These disordered structures have attracted considerable attention for a range of industrial applications including wide-angle displays and paints due to the color's homogeneity and angle-independence. Most of the optical research on disordered structures has focused on assemblies of non-absorbing particles like silica,[6–8] polystyrene (PS),[9–11] and poly(methyl methacrylate) (PMMA)[12] to produce isotropic structural colors like those in birds such as cotingas.[13–16] These studies on disordered structures have found that the particles' form factor, nuances in structural organization, and refractive index (RI) contrasts in multi-component systems are the dominant factors influencing the color generated.[10–12,17–19] In contrast to the many studies on non-absorbing particles, fewer studies have focused on the color production of assemblies comprised of strongly absorbing particles, despite their widespread use in nature and human-made materials. For example, many avian species produce a wide gamut of structural colors ranging from vibrant iridescent[20–22] to earthy colors[23] via self-assembly of broadband-absorbing melanin particles. Among the many multifunctional properties[24] exhibited by melanin due to its chemical and structural diversity,[25,26] melanin possesses two unique optical properties - *high RI* and *broadband absorption*[27,28] which contribute to structural coloration,[29] photoprotection,[30,31] and thermoregulation[32–34] in many biological systems.

Little is known about how strongly absorbing materials like melanin contribute to color production. A few reports show that melanin enhances color saturation/purity by absorbing incoherently scattered light.[35–37] However, experimental work on spherical supraparticles (referred to as supraballs) formed via reverse emulsion-based assembly of synthetic melanin demonstrate that the color can be tuned either by varying the spacing between the melanin particles[38] or the degree of interaction and stratification between melanin and its non-absorbing counterpart (*i.e.*, silica) in the case of binary mixtures.[39] These results[38,39] contradict the idea that melanin solely enhances color saturation/purity and suggest that melanin's structural arrangement can be varied to generate specific hues. These observations highlight the need to quantify the effect of melanin's broadband absorption on structural coloration and to delineate the influence of melanin's design parameters, both in one-component system (degree of absorption, packing order, and size dispersity) and in a binary mixture with a non-absorbing component (the extent of phase separation and stratification), on color generation.

In this work, we perform optical modeling with finite-difference time-domain (FDTD) method[40,41] on one-component supraballs with only silica nanoparticles (for validation purposes as non-absorbing materials have been extensively modeled), one-component supraballs with only synthetic melanin nanoparticles (hereafter referred to as *melanin*), and supraballs composed of binary mixtures of melanin and silica nanoparticles to elucidate the role of melanin in structural color generation. Other models like diffusion theory,[42–44] single-scattering approximation based on Mie scattering[10–12] and Monte-Carlo-based multiple scattering models[45] provide physical insights into the reflectance spectrum, but all of these models are built on the assumptions that a) real systems can be expressed as effective-medium approximations, b) structures are isotropic, and c) near-field couplings are absent. The FDTD algorithm overcomes these assumptions and, in addition, can handle any arbitrary geometry with complex hierarchical structuring, large RI contrasts, and high broadband absorption.

We validate our optical modeling approach by comparing the simulated reflectance spectra for one-component melanin supraballs and one-component silica supraballs with experimental results. We isolate the influence of absorption, packing fraction, and size dispersity on color generation in one-component melanin supraballs in our optical simulations. We also extend our optical model to study how nanoparticle stratification and degree of interparticle mixing of melanin and silica nanoparticles within a supraball affect optical properties. These results provide design principles for synthesizing melanin-based systems to control color, saturation, and lightness factor for applications as pigments for cosmetics, paints, and food coloring.[46,47]

## 2. Results and Discussion

We model color generation using FDTD and coarse-grained molecular dynamics (CG-MD) simulations-generated supraball structures with the CG-MD approach previously validated by direct comparison to experiments.[39,48] The CG-MD simulations represent the melanin and silica nanoparticles (in the case of binary mixtures) as spheres. The nanoparticles are placed under a shrinking spherical confinement to mimic an emulsion-based assembly process[38,39] for supraball generation. We probe the desired design space by varying the interparticle interactions, particle-interface interactions, and particle size dispersity in the CG-MD simulations to generate a diverse set of supraball structures. The resulting particle positions and sizes from the CG-MD simulations are directly input into the FDTD toolbox. The FDTD calculations produce a simulated reflectance spectra by which key color-defining parameters (like primary reflectance peak wavelength, primary reflectance peak width, chromaticity coordinates, saturation, and lightness factor) are obtained allowing for a direct connection between the supraball structure and ensuing optical performance.

## 2.1. Experimental Validation of FDTD Simulations

We validate our optical modeling approach by comparing the simulated reflectance spectra of one-component melanin and one-component silica supraballs with experimentally measured reflectance spectra (**Figure 1**). In the past, researchers have modeled colloidal assemblies of non-absorbing nanoparticles,[45] and this validation is an important step towards understanding the contribution of absorbing melanin nanoparticles in color generation. We need data on material properties (complex RI), size and dispersity of particles, and packing of these nanoparticles for optical simulations. The particle size and dispersity are obtained using small-angle neutron scattering of dilute suspensions of melanin and silica particles (Supporting Information **Figure S1**). We use the complex RI of melanin measured using spectroscopic ellipsometry.[27] We are not able to directly measure the packing fraction in experiments and use the packing fraction of 0.6 as an approximation, in accordance with previous work.[39] **Figure 1** illustrates that the combined CG-MD and FDTD modeling yields a similar reflectance profile as experiments with a close match of the peaks and troughs in simulated and experimental spectra, well within the experimental standard deviation for both the chemistries. Having validated our computational approach, we now apply the optical modeling to explore the importance of all three design parameters: absorption, packing fraction, and size dispersity, on melanin-driven color generation.

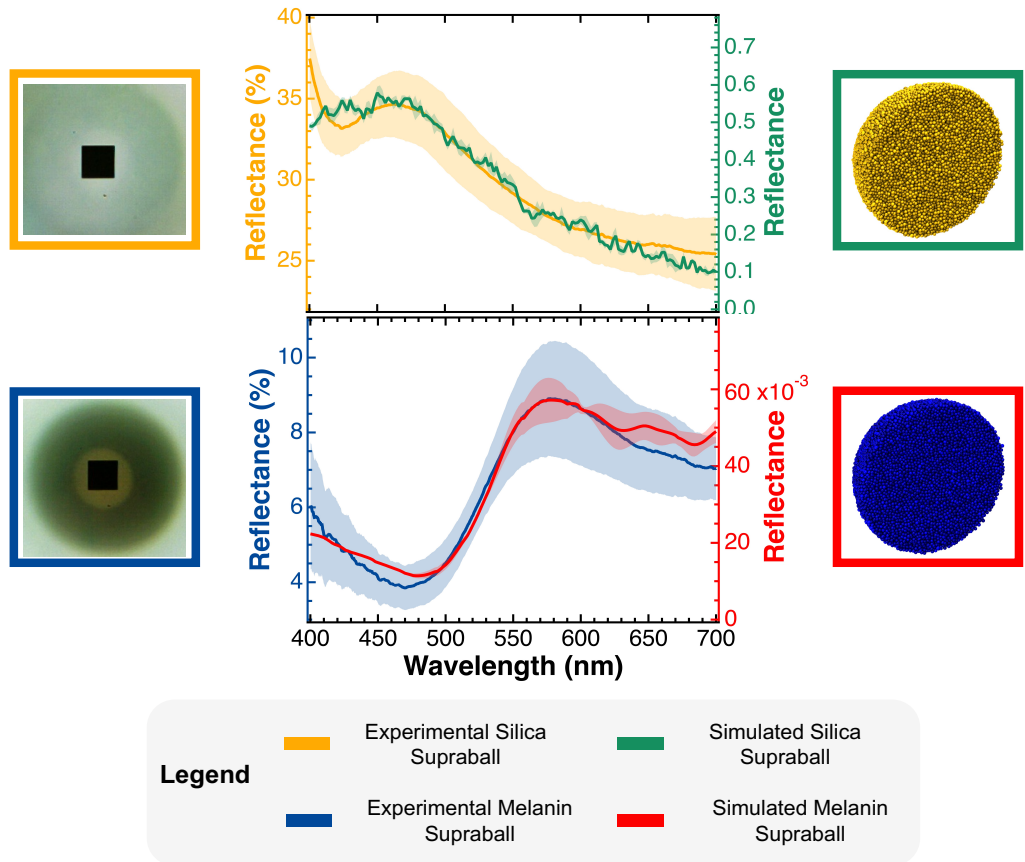

**Figure 1. Experimental validation of computational approach using single component melanin and single component silica supraballs as test samples.** Optical micrographs (*left*) and cross-sections of simulated supraballs (*right*) are shown with the corresponding image border color set to match the legend. The black box in the inset of the optical micrographs represents the size of the area probed during the optical measurements (3 μm x 3 μm). The experimental reflectance curves (yellow and blue) are averaged over 15 supraball samples and the shaded region represents error bars as standard deviation. The simulated reflectance curves (green and red) are averaged over 3 CG-MD simulated supraball structures and the shaded region represents error bars as standard deviation.

## 2.2. Impact of Melanin' Broadband Absorption on Structural Color

Melanin's broadband absorption term is thought to solely enhance color purity.[35–37] Since there has not been any quantitative study on how melanin' broadband absorption affects reflectance properties, we seek to test this hypothesis and delineate the role of its broadband absorption in structural color production. Using FDTD calculations, we systematically vary the magnitude of melanin's absorption term (imaginary part of the complex RI) to outline the impact on the resulting reflectance spectra (**Figure 2a**). While in an experimental approach, one would have to painstakingly synthesize different composite nanoparticles with varying degrees of effective broadband absorption term to perform this study, our computational approach enables us to scale melanin's absorption term from 0% (non-absorbing) to 100% (melanin's typical absorption contribution) keeping the same supraball structure to eliminate any interferences from structural effects. Interestingly, we observe that the primary reflectance peak wavelength exhibits a bathochromic shift (red shift) with increasing absorption contribution (**Figure 2b**; $\beta$ = 0.26 ± 0.03, $R^2$ = 0.81, $p < 0.0001$; **Figure S2a**). We note a decrease in primary reflectance peak width with increasing absorption contribution (**Figure 2c**; $F_{5,12}$ = 34.41, $p < 0.0001$). Even a small input from the absorption term decreases the primary reflectance peak width to nearly half its original value (0% absorbing system), staying constant through higher absorption contributing systems. Additionally, **Figure 2d** confirms previous accounts of melanin increasing color saturation. We find that saturation increases with absorption contribution ($F_{5,12}$ = 479.00, $p < 0.0001$). It is important to note that at ~25% absorption contribution and beyond, the saturation term reaches a plateau (*i.e.*, does not significantly differ) while the lightness factor steadily decreases with increasing absorption (**Figure 2e**; $F_{5,12}$ = 1521.96, $p < 0.0001$). **Figure 2** shows results from supraballs of *monodisperse* melanin nanoparticles. The impact of melanin's broadband absorption term on polydisperse (20% size dispersity) melanin nanoparticle supraballs is reported in **Figure S3**. We observe a similar trend in primary reflectance peak wavelength (**Figure S3b**; $F_{1,4}$ = 60.63, $p$ = 0.0015; **Figure S2b**), primary reflectance peak width (**Figure S3c**; $F_{1,4} = 243.41$, $p < 0.0001$); saturation (**Figure S3d**; $F_{1,4} = 56.87$, $p = 0.0017$), and lightness factor (**Figure S3e**; $F_{1,4} = 4376.40$, $p < 0.0001$) between 0% and 100% absorbing polydisperse nanoparticle supraballs.

These findings bring a new perspective to light where melanin' broadband absorption term not only increases color saturation at the cost of brightness[35] but also produces a red shift in the primary reflectance peak wavelengths, creating an additional parameter to control color. Furthermore, this study also presents an opportunity to secure an effective balance between color saturation and lightness factor. Overall, we emphasize the importance of melanin's broadband absorption term in structural color generation and suggest designing composite materials with low but non-zero absorption contribution (in this case ~25%) to enhance color saturation without significantly comprising brightness.

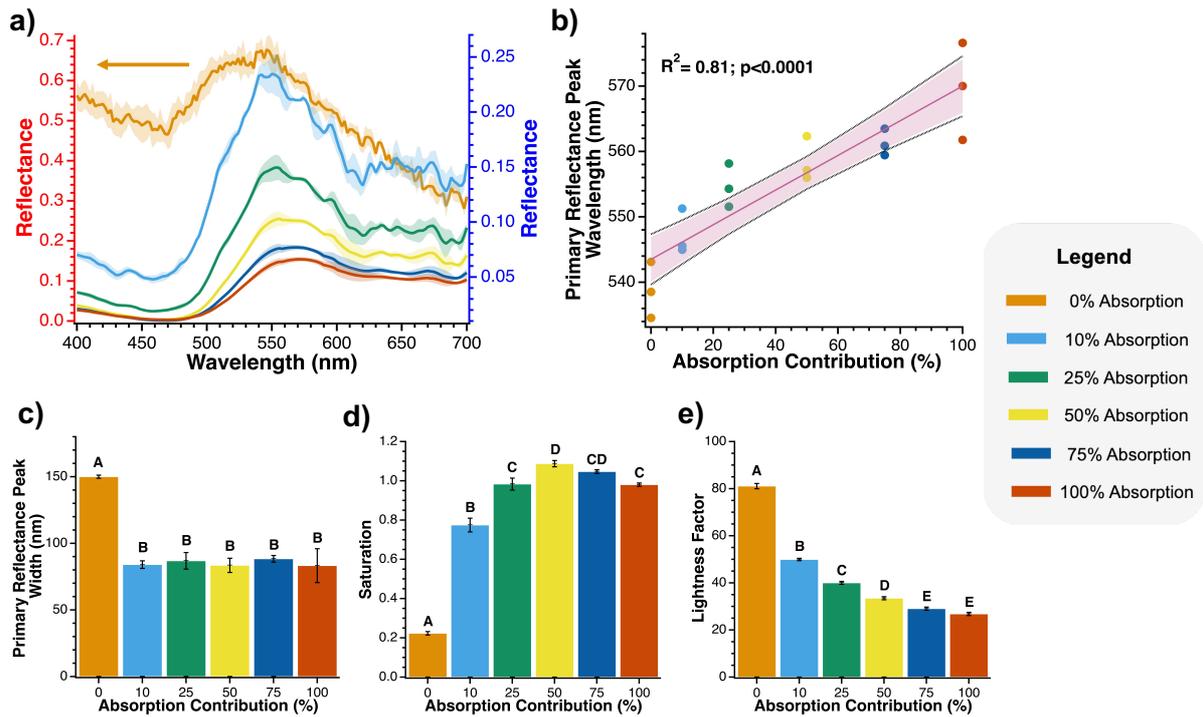

**Figure 2. Effect of varying synthetic melanin's broadband absorption term on structural coloration of monodisperse synthetic melanin supraballs.** (**a**) Average reflectance spectra (*n* = 3 CG-MD simulated supraballs) for each % of melanin's total absorption contribution with the shaded area as the standard deviation. (**b**) The primary reflectance peak wavelength undergoes a red shift with increasing % absorption contribution. The shaded band represents the 95% confidence interval for the linear fit. (**c**), (**d**), and (**e**) show variations in primary reflectance peak width, color saturation, and lightness factor, as a function of % absorption contribution, respectively, with error bars as the standard deviation. We ran a Tukey HSD test to perform multiple pairwise comparisons between the means of the groups. For categories that do not differ significantly ($p > 0.05$) the same letters are provided.

## 2.3. Effect of Melanin Packing Fraction on the Supraball Reflectance

After clarifying the importance of melanin's broadband absorption on structural color, we consider melanin packing fraction as a possible design parameter to tune supraball reflectance. In **Figure 3**, we explore how adjusting the packing fraction ($\varphi$) of monodisperse melanin particles inside the supraball affects the structural color. The packing fraction of the one-component melanin supraball is decreased after the formation of the close-packed supraball ($\varphi \approx 0.6$) by randomly removing nanoparticles, similar to experiments that use chemical etching to generate voids.[49,50] Thus, we study how changes to the melanin supraball structure impact the structural color without adjusting the melanin nanoparticle characteristics. Cross-sections of the simulated supraballs used in the FDTD calculations are provided in **Figure 3a**, and we also consider a hexagonally close-packed (HCP) crystalline supraball for comparison to our disordered (amorphous) supraballs. Experimentally, one could produce an HCP supraball by adjusting the assembly time and nanoparticle dispersity.[51] **Figure 3b** shows the reflectance spectra for all systems, and we note that the disordered supraballs exhibit a peak reflectance value that is appreciably lower than the peak reflectance value for the HCP supraball. As expected of any crystalline photonic assembly, the HCP supraball has a narrow reflectance peak compared to the disordered supraballs with enhanced saturation and brightness. **Figure S4a-b** shows that the primary reflectance peak wavelength and width consistently decrease with increasing packing fraction ($F_{2,6} = 45.25$, $p = 0.0002$; $F_{2,6} = 18.73$, $p = 0.0026$) for disordered supraballs. While examining the CIE 1931 chromaticity diagram in **Figure 3c**, we find that reducing the packing fraction of the supraball has minimal impact on the spectral colors while the crystalline HCP supraball traverses a distinct distance along the color space relative to the disordered supraballs of varying packing fraction.

As we are primarily interested in understanding the color generation from disordered supraballs, **Figure 3d-f** focuses on the supraballs with varying packing fraction, excluding the HCP crystalline supraball. We find that increasing the packing fraction increases the primary structure factor [S(q)] peak height as one might expect from increasing the number of particles in the supraball, but the primary S(q) peak width (full width at half maximum (FWHM), $\Delta q$) does not vary with packing fraction (**Figure 3d**). Moreover, since we are dealing with monodisperse particles, the primary S(q) peak position ($q_{pk}$) also remains constant for all cases. **Figure 3e** illustrates that the color saturation significantly increases with increasing S(q) peak height ($\beta = 0.18 \pm 0.02$, $R^2 = 0.89$, $p < 0.0001$) while **Figure 3f** suggests the lightness factor decreases minimally (statistically) with S(q) peak height ($\beta = -1.96 \pm 0.92$, $R^2 = 0.39$, $p = 0.0702$). **Figure S4c-d** shows that color saturation and lightness factor do not vary with the $\Delta q$ (illustrated as $q_{pk}/\Delta q$ in the plots) as that tends to be consistent for all packing fractions considered ($\beta = 0.004 \pm 0.1$, $R^2 = 0.0002$, $p = 0.9680$; $\beta = 0.90 \pm 1.61$, $R^2 = 0.04$, $p = 0.5926$).

Overall, **Figure 3** suggests that adjusting the melanin packing fraction minimally impacts structural color and reveals a minor increment in color saturation at a trifling cost on brightness. This again confirms that using a packing fraction of 0.6 to compare optical simulation data with experiments in **Figure 1** was reasonable. Because the primary peak reflectance wavelengths and peak widths vary across the different cases of supraball packing fraction, the total contribution of the reflectance spectra yields marginally distinguishable colors with the color differences being highest for the two ends of the disordered packing fractions ($\Delta E_{0.4-0.6}$ = ~3.3 times the "just noticeable difference" (JND) value).[52]

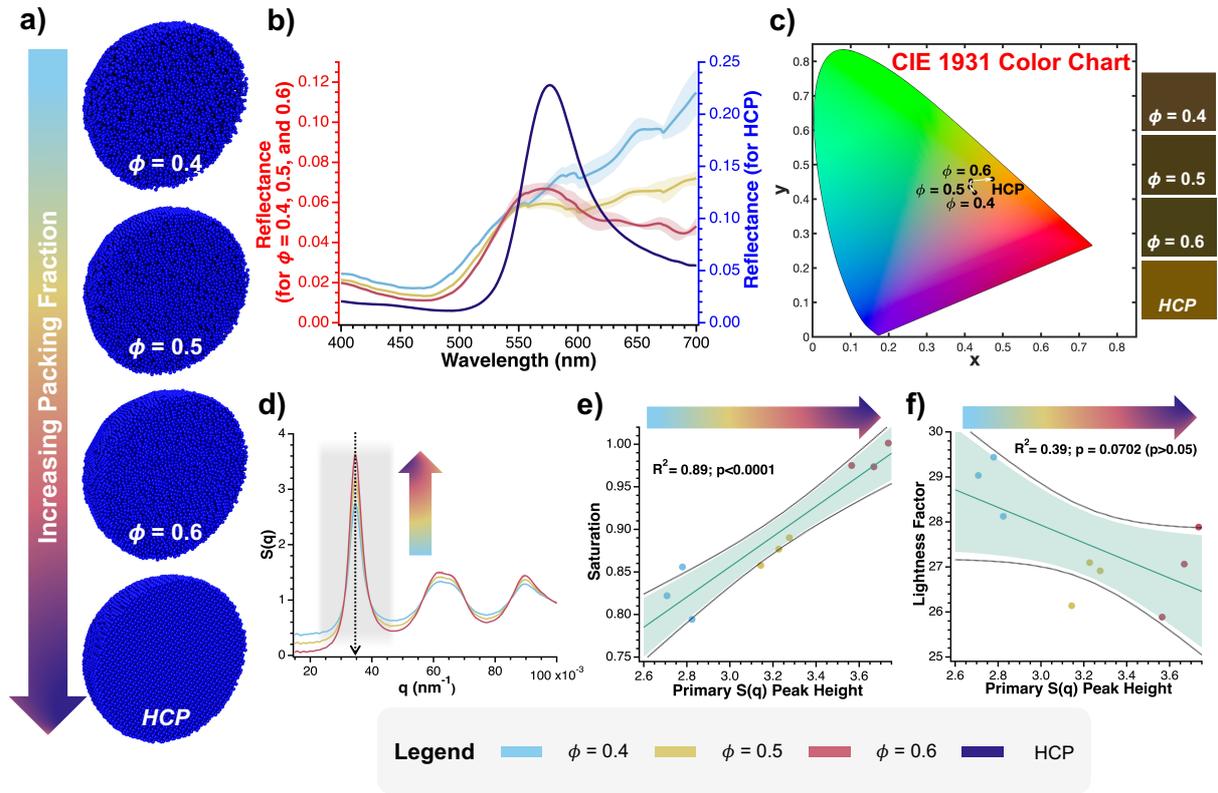

**Figure 3. Effect of melanin packing fraction ($\varphi$) on the supraball reflectance.** (**a**) Visualizations of the cross-section of supraballs of varying packing fractions arranged in the increasing order from top to bottom. (**b**) Average reflectance spectra ($n$ = 3 CG-MD simulated supraballs) for each case of packing fraction with the shaded area as the standard deviation. (**c**) Color changes with increasing packing fraction are represented in the chromaticity diagram (CIE 1931 color chart) and as RGB color panels. (**d**) Supraball structure factors for different packing fractions (short-range ordered) with focus on primary structure factor peak (shaded in gray). The primary structure factor peak height increases with packing fraction. The color saturation increases with primary structure factor peak height (**e**) while there is an insignificant change in brightness (**f**). The shaded bands in (e) and (f) represent the 95% confidence interval for the linear fit.

**2.4. Effect of Melanin Size Dispersity on the Supraball Reflectance**

Relative to melanin packing fraction, melanin nanoparticle size dispersity has a larger effect on the supraball reflectance. **Figure 4** provides a detailed exploration of the particle size dispersity effect on the structural coloration of supraballs. The cross-sections of the simulated supraballs under investigation along with the lognormal nanoparticle size distributions are provided in **Figure 4a**. The supraball reflectance spectra exhibits a non-monotonic behavior with size dispersity such that a small increase in dispersity from monodisperse to ~1% increases the peak reflectance while larger dispersities show consistent decrease in the peak reflectance values (**Figure 4b**). Interestingly, the reflectance spectra suggest that increasing size dispersity causes a red shift of the primary reflectance peak wavelength. However, at higher dispersities, secondary peaks at lower wavelengths become prominent, as illustrated for the 20% dispersity case, pushing the original primary peak to higher wavelengths beyond the visible spectrum. **Figure S5a** confirms the non-monotonic behavior of the primary reflectance peak wavelength with size dispersity ($F_{4,10} = 62.13$, $p < 0.0001$). **Figure 4c** demonstrates this trend with 0% to 10% dispersity steadily producing structural colors with an increasing red contribution, while the 20% dispersity traverses to a greenish-blue region with a darker shade (turning nearly black). Interestingly, this resembles an observation in nature where deep-sea dragon fish generate their ultra-dark blackish-blue skin color from highly polydisperse melanosomes (melanin-containing organelles) in their skin.[53] This supports our finding that large melanin polydispersity ultimately results in a hypsochromic shift (blue shift) to very dark shades of structural colors. **Figure S5b** shows that the primary reflectance peak width gradually grows with larger polydispersity ($F_{4,10} = 7.13$, $p = 0.0055$). To further connect the optical responses to the supraball structure, we examine the melanin $S(q)$ in **Figure 4d**. We note that the primary $S(q)$ peak height decreases and the $\Delta q$ peak width increases as is typically seen as size dispersity increases.[54] Here, since the different polydisperse supraball structures share the same lognormal mean particle diameter, the $q_{pk}$ remains constant. **Figures 4e** and **4f** connect the $\Delta q$ to the color saturation and lightness factor obtained for each supraball system. The saturation and lightness factor significantly decrease as the $\Delta q$ increases ($q_{pk}/\Delta q$ decreases): saturation: $\beta = 0.19 \pm 0.01$, $R^2 = 0.97$, $p < 0.0001$ and lightness factor: $\beta = 2.42 \pm 0.30$, $R^2 = 0.83$, $p < 0.0001$. We see a similar trend when correlating primary S(q) peak height changes to saturation ($\beta = 0.41 \pm 0.02$, $R^2 = 0.97$, $p < 0.0001$) and lightness factor ($\beta = 5.15 \pm 0.72$, $R^2 = 0.80$, $p < 0.0001$), as shown in **Figure S5c-d**.

In the more polydisperse samples, different sizes segregate within the self-assembled supraball structures.[39,48] The Young-Laplace equation states that the nanoparticle-interface attraction scales with the nanoparticle cross-section[55,56] resulting in an enrichment of larger diameter nanoparticles on the supraball surface, particularly for high dispersity systems as shown in (**Figure S6a**). We quantify the effect of a 20% dispersity melanin supraball formed under a non-attractive interface that prevents the size-based stratification of melanin nanoparticles on the supraball surface to illustrate a non-intuitive approach to tune the structural color (**Figure S6**). We note that the reflectance spectra for the cases of an attractive and non-attractive interface possesses distinct shapes, and the non-attractive interface supraball does not have a reflectance peak centered around 500 nm (**Figure S6b**). **Figure S6c** shows that the resulting structural color exhibits a strong red shift moving from an attractive interface to a non-attractive one. In **Figure S6d-f**, we find that the primary reflectance peak wavelength, color saturation, and lightness factor all significantly increase for supraballs formed under a non-attractive interface ($F_{1,4} = 903.57$, $p < 0.0001$; $F_{1,4} = 43.37$, $p = 0.0028$; $F_{1,4} = 151.14$, $p = 0.0003$). Thus, controlling size-based stratification in a polydisperse one-component supraball provides for additional fine-tuned control over the resulting structural color, saturation, and lightness factor.

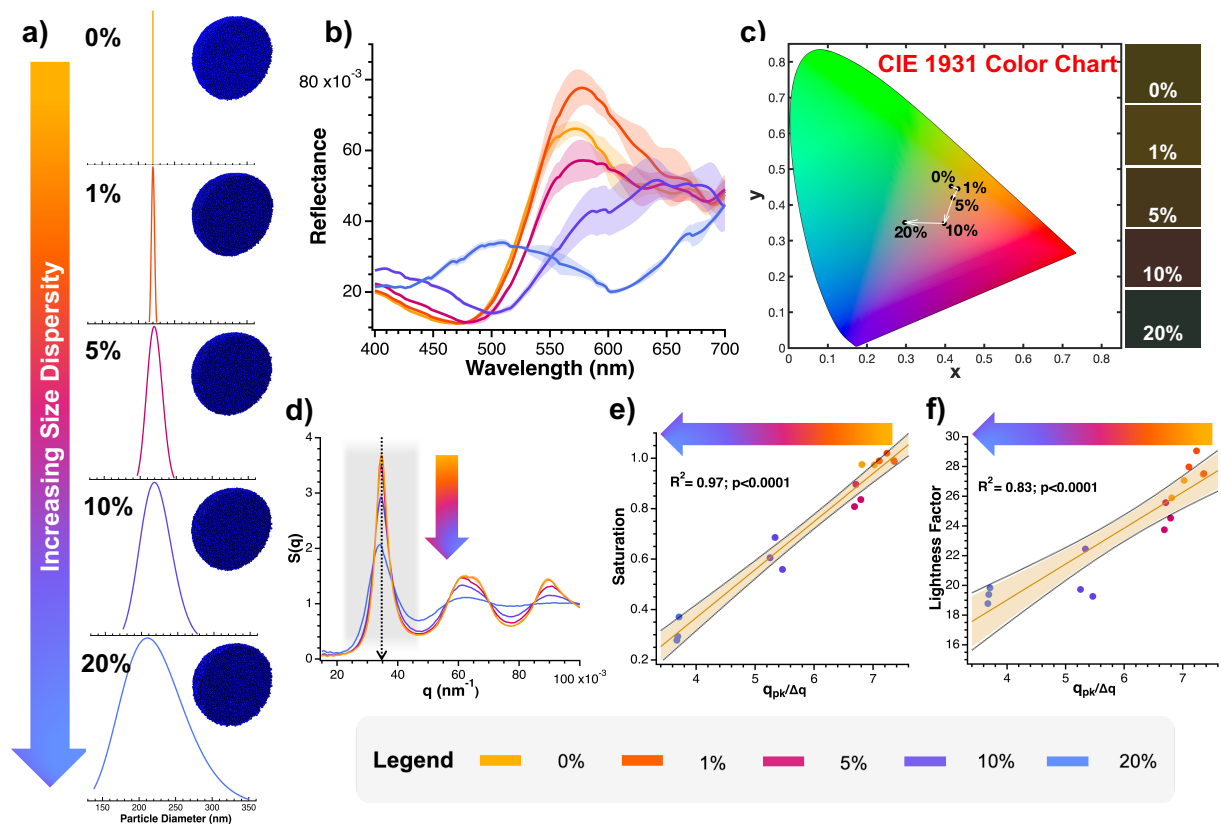

**Figure 4. Impact of melanin nanoparticle size dispersity on the supraball reflectance.** (**a**) Lognormal size distributions with visualizations of the cross-section of supraballs of varying size dispersities (inset) arranged in the increasing order from top to bottom. (**b**) Average reflectance spectra (*n* = 3 CG-MD simulated supraballs) for each case of size dispersity with the shaded area as the standard deviation. (**c**) Color changes with increasing size dispersity are represented in the chromaticity diagram (CIE 1931 color chart) and as RGB color panels. (**d**) Supraball structure factors for different size dispersity cases with focus on primary structure factor peak (shaded in gray). The primary structure factor peak height increases, and the peak width decreases with size dispersity. The color saturation (**e**) and brightness (**f**) decrease with primary structure factor peak width. The shaded bands in (e) and (f) represent the 95% confidence interval for the linear fit.

**2.5. Effect of Degree of Mixing of Absorbing and Non-absorbing Particle Mixtures on the Supraball Reflectance**

Up until now, we have focused on one-component supraballs solely created from melanin nanoparticles. In the next two sections, we extend our study to two-component supraballs produced from a mixture of both absorbing and non-absorbing components. Many avian species generate structural colors by strategically arranging the broadband absorbing melanosomes within a non-absorbing keratin matrix to yield either iridescent or non-iridescent earthy colors.[20,23] In the case of disordered assemblies of melanosomes that yield colors in feathers, the structures range from weakly segregated (such as in ravens) to strongly segregated in large domains (as in California quail).[57] Researchers have used this strategy of mixing absorbing and non-absorbing components to tune structural colors in the past.[39,58] However, whether the extent of mixing between the absorbing and non-absorbing species affects structural coloration, as has been observed in case of some avian species, has not been systematically studied yet.

In **Figure 5**, we explore a wide range of two-component mixing in the supraballs from strongly mixed to strongly segregated (demixed) phases by adjusting the interaction between the melanin and non-absorbing particles (in this case silica) to generate the desired amount of phase mixing. Experimentally, one can adjust the effective nanoparticle interactions to achieve specific mixing by modifying the surface chemistry of either or both of the two particle types via surface functionalization with small molecules or functional polymer grafts.[59,60] For visual reference, the supraballs with varying degrees of mixing are shown in **Figure 5a** with melanin represented as blue spheres and silica as yellow spheres. Surprisingly, **Figure 5b** reveals that all types of nanoparticle mixing, besides strongly demixed cases, have negligibly different reflectance spectra with similar primary reflectance peak positions, peak heights, and peak widths. The strongly demixed case shows a reduced reflectance behavior with strong contributions from the lower wavelength regions. These combined effects produce a dark greyish color unlike all other mixing states that exhibit a green color and are crowded at a similar region of the color chart (**Figure 5c**). We can extract structural information (partial S(q)s) from these morphologies (**Figure S7**) to learn about the strength of interaction between melanin and silica particles. The cross-correlation term, given by $S(q)_{Mel-Sil}$ in **Figure 5d**, tells us that the higher the partial S(q) peak height, the stronger the interaction between the two components and hence the greater the degree of mixing. **Figure 5d** distinctly shows that the primary partial S(q) peak height for the strongly demixed case is appreciably different (lower) than the rest of the mixing cases. This result is in line with the reflectance behavior in **Figure 5b** and the colors produced in **Figure 5c**. Furthermore, **Figure 5e** and **5f** show that color saturation and lightness factor vary with degree of mixing ($F_{4,10} = 12.59$, $p = 0.0006$; $F_{4,10} = 11.77$, $p = 0.0008$), and the *post-hoc* tests confirm the consistency in the trends observed previously where the strongly demixed case is significantly different than rest of the groups. It is interesting to note that a critical threshold of phase separation must be met to notice distinguishable changes in the reflectance properties of binary mixed systems. In this study, we consider equal composition of melanin and silica in the binary mixture supraball systems. Future studies will entail investigating this range of mixing states across different compositions of melanin and silica to understand compositional sensitivity to phase separation and optical behavior and delineate the optimal conditions for obtaining brighter and saturated colors.

Up to this point, we have looked at the effect of degree of mixing between monodisperse melanin and silica nanoparticles within a supraball geometry. Avian species displaying colors, like California quail, produce polydisperse melanosomes that are mixed in the keratin matrix.[57] Furthermore, in Section 2.4, we find that melanin nanoparticle size dispersity has a large influence on structural colors. Hence, we seek to understand how size dispersity will impact the optical reflectance for a given mixed state. To address this question, we look at the binary mixture supraball morphologies representing the two extreme ends of mixing states (strongly demixed and strongly mixed) with nanoparticle size dispersities varying as 0%, 10%, and 20% (**Figure S8a**). The simulated optical reflectance obtained from these structures (**Figure S8b**) show that the size dispersity strongly diminishes the primary reflectance peak resolution with the emergence of dull dark colors, irrespective of the mixing state of the particles. While the monodisperse strongly demixed and strongly mixed systems produce visible color differences (~9 times the JND), the strongly demixed and strongly mixed supraballs with 10% or higher dispersity produce colors with no visually observed color difference. **Figure S8c** shows that color saturation decreases with size dispersity for the strongly mixed state ($F_{2,6}$ = 125.09, $p <$ 0.0001) and varies negligibly with size dispersity for the strongly demixed state ($F_{2,6}$ = 1.66, $p$ = 0.2667). The lightness factor also follows the same trend for both mixing degrees wherein the lightness factor decreases with increasing size dispersity (**Figure S8d**; strongly demixed: $F_{2,6}$ = 6.88, $p$ = 0.0280, strongly mixed: $F_{2,6}$ = 74.93, $p <$ 0.0001). In conclusion, for the systems considered, the degree of mixing of equal composition melanin and silica supraballs is not a design parameter that strongly influences the structural coloration because the nanoparticle size dispersity dictates the optical response.

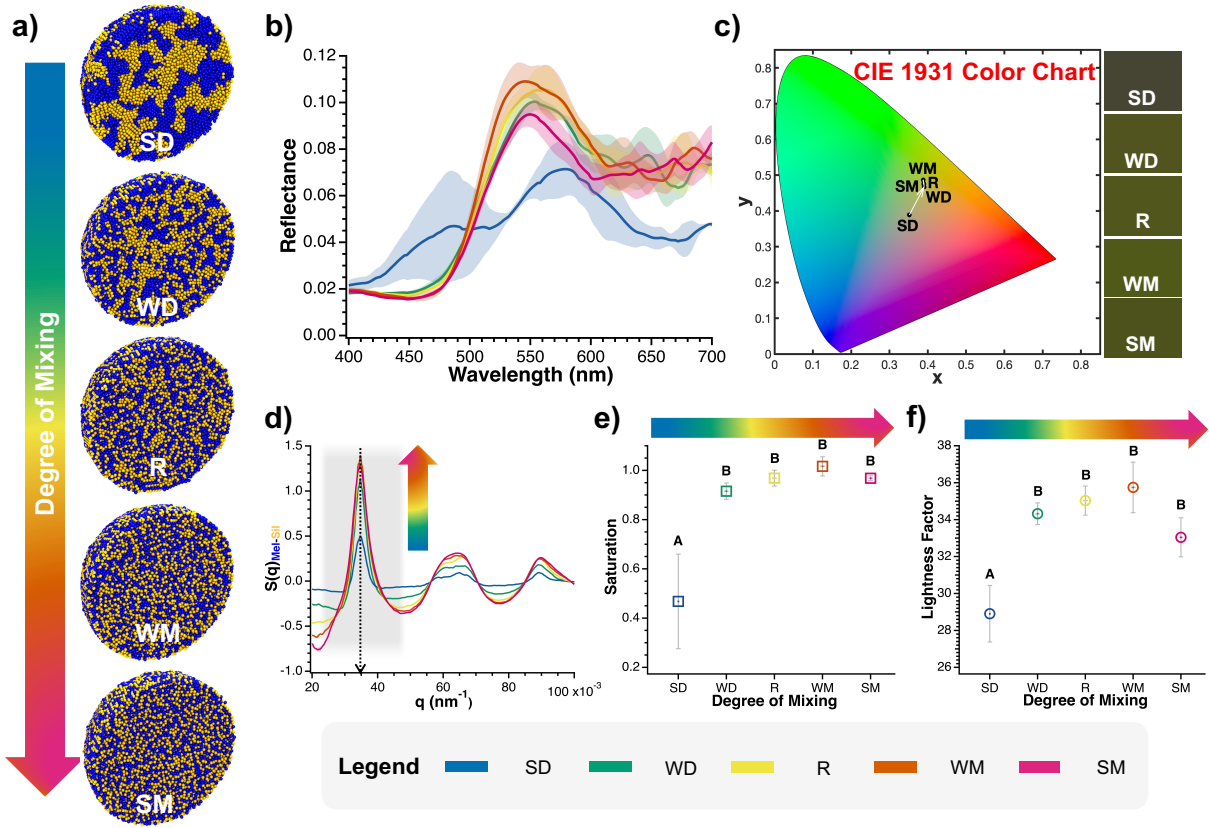

**Figure 5. Influence of degree of mixing of melanin and silica particle mixtures on the supraball reflectance.** (**a**) Visualizations of the cross-section of supraballs with varying levels of particle mixing arranged in the increasing order from top to bottom (SD: strongly demixed; WD: weakly demixed; R: randomly mixed; WM: weakly mixed; SM: strongly mixed). (**b**) Average reflectance spectra ($n = 3$ CG-MD simulated supraballs) for each type of mixing state with the shaded area as the standard deviation. (**c**) Color changes with increasing order of particle mixing are represented in the chromaticity diagram (CIE 1931 color chart) and as RGB color panels. (**d**) The binary mixture supraball partial structure factor (cross-correlation term) for different mixing cases with focus on primary partial structure factor peak (shaded in gray). The primary partial structure factor peak height increases with level of mixing. (**e**) and (**f**) show variations in color saturation and lightness factor as a function of degree of mixing, respectively, with error bars set as the standard deviation. We ran a Games-Howell non-parametric *post-hoc* test for color saturation and a Tukey HSD parametric *post-hoc* test for lightness factor for multiple pairwise comparisons between means of the groups. For categories that do not differ significantly ($p > 0.05$), the same letters are provided.

**2.6. Effect of Absorbing/Non-absorbing Shell Formation on Supraball Reflectance**

Earlier experimental work with melanin and silica binary mixture supraballs discovered, during the reverse-emulsion assembly process, that melanin nanoparticles formed a shell on the supraball surface likely impacting the resulting structural colors.[39] Hence, we conclude our investigation by systematically examining the impact of nanoparticle shell formation where one type of nanoparticle enriches the supraball surface (**Figure 6**). In an experimental system, one could accomplish a shell formation by adjusting the nanoparticle-interface contact angles[39] and interaction strengths[48] or the rate of emulsion assembly for different sized components.[61] We also consider a significantly more complex case of a completely stratified core-shell supraball where all the nanoparticles of one type are located at the supraball center (core) while all the other type of nanoparticles form a shell. Experimentally, these core-shell supraballs could be realized by using a double-emulsion or two-step emulsion assembly process to first form a core of one particle type and then add a shell of the second particle type.[62,63] **Figure 6a** illustrates the systems considered from a core-melanin shell-silica supraball to a core-silica shell-melanin supraball expressing different degrees of stratification. For the intermediate systems that are not fully stratified, we consider a random degree of mixing. We expect from **Figure 5** that the results for random mixing will hold for other cases besides the strongly demixed system. **Figure S9** quantifies the degree of stratification for all supraballs by plotting the average silica composition as a function of radial distance from the supraball center to the supraball surface. We find that varying the degree of stratification significantly impacts the resulting reflectance spectra (**Figure 6b**) and the resulting structural color (**Figure 6c**) that spans from blue to green to orange-red. In fact, the degree of stratification produces the most diverse colors of all design parameters considered in this study. In **Figure 6d**, we find that the primary reflectance peak wavelength significantly increases ($F_{4,10} = 301.17$, $p < 0.0001$) moving from a core-melanin shell-silica supraball to a core-silica shell-melanin supraball. Interestingly, the melanin shell supraball possesses a higher contribution from red wavelengths that dials down the contribution from yellow wavelengths (**Figure 6b**) yielding darker shades. **Figure 6e** demonstrates that the color saturation significantly varies based on the degree of stratification ($F_{4,10} = 61.63$, $p < 0.0001$) with the two core-shell supraballs having the minima and maxima respectively. We find that going from the core-melanin shell-silica to silica shell supraball produces a larger change in saturation compared to moving from the melanin shell to core-silica shell-melanin supraball. The lightness factor also varies with degree of stratification ($F_{4,10} = 26.33$, $p < 0.0001$) via a non-monotonic trend with silica shells possessing high lightness factors, the no shell and core-silica shell-melanin obtaining intermediate lightness factor values, and the melanin shell exhibiting the lowest lightness factor (**Figure 6f**). This trend is nonintuitive as one would expect a thicker broadband absorbing melanin shell (in the core-silica shell-melanin supraball) to produce the lowest brightness.

The results show a unique way to tune structural colors simply using two components by controlling the extent of stratification. These observations not only present an opportunity for industrial design of colloidal assemblies to produce tunable structural colors but also sheds light on nature's design principles. For example, in avian species, having a dense basal layer of melanin[64,65] or a thin layer of non-absorbing cortex (*i.e.* keratin) in feather barbules[57] can yield brighter structural colors with saturation, as has also been reported in our computational study for the two silica shell morphologies. Furthermore, our results on melanin shell segregation enable us to endorse previously reported observations[39] that segregated melanin shell improves color saturation. However, this comes at a cost of brightness which one can solve by ensuring complete segregation of the particles to the interface with an underlying non-absorbing component.

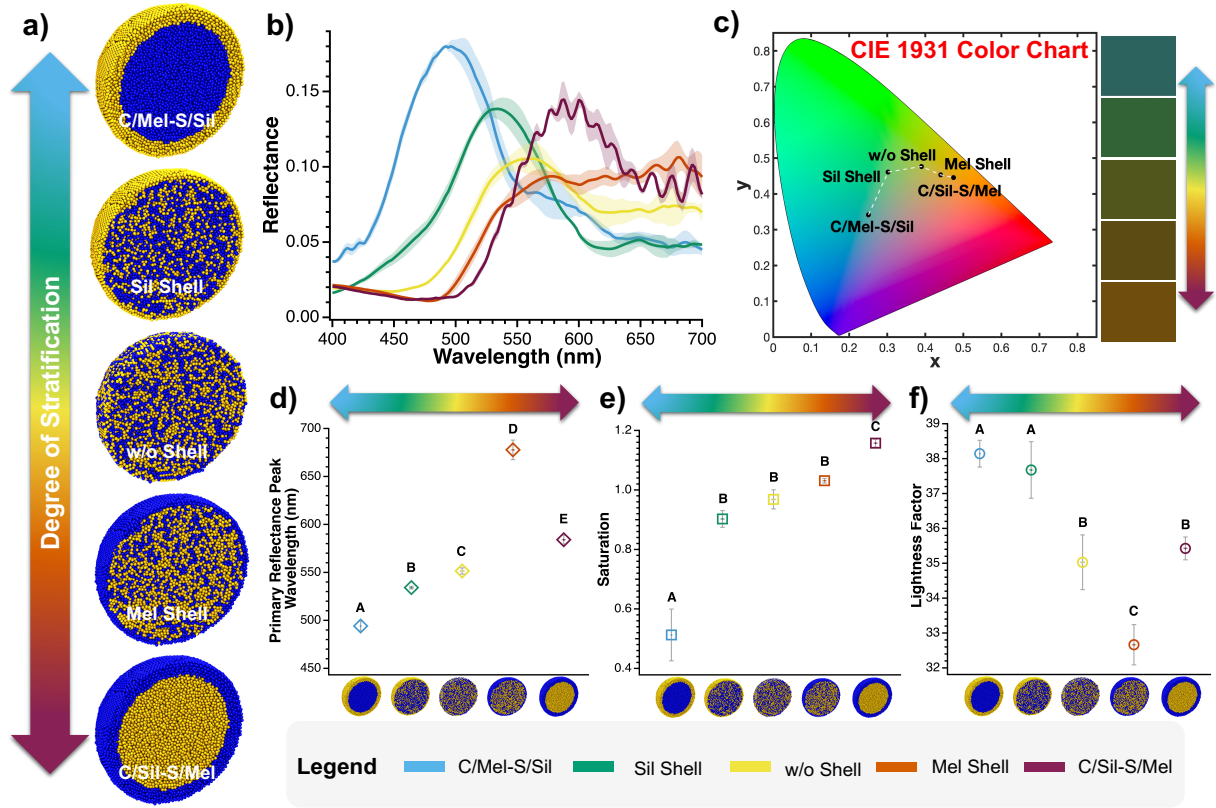

**Figure 6. Effect of silica/melanin shell stratification on the supraball reflectance.** (**a**) Visualizations of the cross-section of supraballs with varying degrees of stratification. (**b**) Average reflectance spectra (*n* = 3 CG-MD simulated supraballs) for each type of shell stratification with the shaded area as the standard deviation. (**c**) Color changes with varying degrees of stratification are represented in the chromaticity diagram (CIE 1931 color chart) and as RGB color panels. (**d**), (**e**) and (**f**) show variations in primary reflectance peak wavelength, color saturation and lightness factor as a function of degree of stratification, respectively, with error bars as the standard deviation. We ran a Games-Howell non-parametric *post-hoc* test for primary reflectance peak wavelength and color saturation, and a Tukey HSD parametric *post-hoc* test for lightness factor for multiple pairwise comparisons between means of the groups. For categories that do not differ significantly ($p > 0.05$) the same letters are provided.

## 3. Conclusion

In this paper, we use a combination of CG-MD and FDTD approaches to model structural color production of self-assembled supraballs of absorbing and non-absorbing nanoparticles. We observe a close match between experiments and our computational approach for single component melanin and silica supraballs. We then use this computational method to compare the influence of absorption, dispersity, and packing fraction on color generation. We observe that strongly absorbing species such as melanin not only improve color saturation at the cost of brightness but also cause a significant red shift in the reflectance peak. Decreasing the melanin-based supraball packing fraction (from 0.6 to 0.4) has minimal impact on the structural color and brightness while significantly reducing the color saturation. Interestingly, we find that increasing the nanoparticle size dispersity from monodisperse to 1% results in a significant increase in color saturation and lightness factor, which was unexpected because a larger increase in dispersity results in a significant decrease in both. Furthermore, we find that changing the nanoparticle interface interaction from attractive to repulsive allows for a significant red shift and dramatic increase in both the color saturation and the lightness factor.

After analyzing one-component melanin supraballs, we expand our study to incorporate binary mixtures of highly absorbing melanin and non-absorbing silica nanoparticles. For the systems we consider, we note that adjusting the degree of mixing has a negligible effect unless the components are strongly demixed suggesting that adjusting the silica-melanin interaction does not provide a suitable method to tune structural colors. Finally, we explore stratifying the two components to produce structures ranging from core-shell to no shell and describe how the degree of stratification spans the largest range of colors produced. These results provide insights into the use of this combined CG-MD and FDTD computational approach to predict colors for conditions that are not yet explored in experiments and offer guidance for designing new structures to tune colors in different regions of the electromagnetic spectra such as ultra-violet (UV) and infrared (IR).

## 4. Methods

This section describes both aspects of our experimental and computational approach: synthesizing nanoparticles, producing self-assembly and characterization; generating supraball structures using coarse-grained molecular dynamics (CG-MD) simulations, and performing finite-difference time domain (FDTD) simulations.

### 4.1 Experimental Method to Produce Supraballs and Characterization

*4.1.1 Particle Synthesis and Characterization*

Silica nanoparticles (SP) were synthesized using a modified Stöber process.[66] The synthetic melanin nanoparticles (SMP) were synthesized following our previous protocol[39] via the auto-oxidation and polymerization of dopamine monomer (Sigma Aldrich), a widely accepted synthetic mimic of natural melanin moiety, in basic environment *i.e.*, in the mixture of water, ethanol, and ammonia solution ($NH_4OH$; Sigma Aldrich - 28 to 30 wt%) at room temperature under constant stirring. To examine the particle morphology, we drop-casted the SP and SMP onto carbon-coated copper grids (FCF200-Cu; Electron Microscopy Sciences) for transmission electron microscopy (JEM-1230, JEOL Ltd.).

To obtain the ensemble average of particles in terms of shape, size, and polydispersity for the experimental validation of optical simulations, we performed small-angle neutron scattering (SANS) on dilute aqueous suspensions of SP and SMP at the National Institute of Standards and Technology Center for Neutron Research (NIST CNR or NCNR). The standard configurations of the beamlines were used to run the SANS experiments *i.e.*, a) for NG7 SANS , 1 m, 4 m, and 13 m sample-to-detector distances were used with 6 Å neutrons while the Lens setup for low-*q* used 8 Å at 15.3 m, and b) for vSANS, the high-*q* setup used 6 Å neutrons with front and middle detector carriages set at 1.1 m and 5.1 m respectively, from the sample while the low-*q* setup used 11 Å neutrons with front and middle detector carriages set at 4.6 m and 18.6 m respectively, from the sample. The dilute suspensions (prepared in deuterated water to avoid incoherent scattering) were contained in quartz banjo cells (Product # 120-2mm; Hellma USA) to avoid any undesired scattering contribution from the containers. The SANS experiments were conducted at ambient temperature and the measured intensities were corrected for background scattering and empty cell contributions. They were also normalized using a reference scattering intensity of a polymer sample of known cross-section. The reduction of raw SANS data was performed following a well-known protocol described by S. R. Kline.[67] The processed SANS data were analyzed using SasView 4.2.2 (https://www.sasview.org). A spherical form factor model for lognormally distributed polydisperse spheres was used to fit the data (**Figure S1**).

*4.1.2 Supraball Preparation and Reflectance Measurement*

We referred to our previous reverse-emulsion assembly protocol to make one-component silica and melanin supraballs.[38,39] This process involved two steps. First, the surfaces of the glass vials used in the assembly process were rendered hydrophobic by growing *n*-octadecyltrichlorosilane (OTS) self-assembled monolayer (SAM) following a modified protocol.[68] To briefly describe this coating process, we added 2 volume% OTS-toluene solution in base-bath cleaned and dried glass vials and degassed the solution using inert compressed nitrogen gas for 16 hours under sealed conditions at room temperature. Later, we washed these vials three times each with toluene and ethanol via ultrasonication and then annealed them at 120°C under vacuum for 2 hours. We inspected the quality of the SAM coating by performing contact angle measurements (water contact angle ~ 112°). This step ensured the dispersed phase (aqueous droplets) of the reverse-emulsion did not adhere to the glass walls and break upon contact. The second step involved preparing the emulsion wherein, typically, 30 μl of aqueous solution of SP/SMP (30 mg/mL) was added to 1 mL of anhydrous 1-octanol (Sigma-Aldrich) and the mixture was vortexed at a speed of 1600 rpm for 2 min followed by 1000 rpm for 3 min to form the reverse-emulsion. The supraballs, thus formed, were allowed to precipitate, the supernatant was extracted, and the samples were dried at 60°C.

The dried supraballs were placed on Piranha-cleaned silicon wafers (substrate) and measured for their reflectance property using a CRAIC AX10 microspectrophotometer (CRAIC Technologies, Inc.). We employed a 50x objective and a 75-W Xenon short arc lamp (Ushio UXL-75XE) for the white light source. Silver mirror was used as the reflectance standard for calibration purposes. The experimental reflectance spectra reported was an average of 15 measurements with the error bars represented by the standard deviation (**Figure 1**).

**4.2 CG-MD Simulations to Produce the Supraball Structures**

The procedure to generate the *in silico* supraball structures was presented in detail in previous work.[39,48] A brief description of the model and method is provided below.

*4.2.1 Model*

The CG model represented the synthetic melanin particles and, for binary systems, silica particles as spheres with an average diameter of 220 *nm*. To incorporate nanoparticle size dispersity, each nanoparticle type was composed of 11 distinct groups of differing diameter spheres with the diameter for each group drawn from a lognormal distribution with the stated standard deviation (represented by the histograms in **Figure S1b**). The smallest diameter group was set at the 1% probability density, and the largest diameter group was set at the 99% probability density. The intermediate groups were uniformly distributed between the largest and smallest groups. For all binary systems, the volume fraction of the melanin nanoparticles was set to ~ 0.5. The characteristic simulation length $\sigma$ was 1.0 nm, and the characteristic mass $m$ was the mass of the smallest melanin nanoparticle determined using the volume and mass density. All other nanoparticles' mass were scaled based on their relative volume and mass density difference between melanin (~ 1.3 g/cm$^3$) and silica (~2.3 g/cm$^3$).[39] All nanoparticles interacted through the colloid Lennard-Jones (cLJ) potential[69] with Hamaker constants set to achieve different particle mixing: randomly mixed (R) supraballs ($A_{Mel-Mel}$ = 0.25 k$_B$T, $A_{Mel-Sil}$ = 0.25 k$_B$T, and $A_{Sil-Sil}$ = 0.25 k$_B$T), weakly demixed (WD) supraballs ($A_{Mel-Mel}$ = 0.25 k$_B$T, $A_{Mel-Sil}$ = 0.20 k$_B$T, and $A_{Sil-Sil}$ = 0.25 k$_B$T), weakly mixed (WM) supraballs ($A_{Mel-Mel}$ = 0.20 k$_B$T, $A_{Mel-Sil}$ = 0.25 k$_B$T, and $A_{Sil-Sil}$ = 0.20 k$_B$T), strongly demixed (SD) supraballs ($A_{Mel-Mel}$ = 0.50 k$_B$T, $A_{Mel-Sil}$ = 0.08 k$_B$T, and $A_{Sil-Sil}$ = 0.50 k$_B$T), and strongly mixed (SM) supraballs ($A_{Mel-Mel}$ = 0.08 k$_B$T, $A_{Mel-Sil}$ = 0.50 k$_B$T, and $A_{Sil-Sil}$ = 0.08 k$_B$T). In all systems, the characteristic energy $\varepsilon$ was 1.0 k$_B$T. To form the spherical supraballs, a spherical wall enforced the spherical confinement by interacting with the nanoparticles. For most systems, the spherical wall interacted with the nanoparticles through an attractive harmonic potential with strong nanoparticle-wall interface strength ($\varepsilon_{attractive\ wall}$ = 500.0 k$_B$T), a cutoff distance set to the nanoparticles' radius (ensuring the potential only impacted nanoparticles at the interface), and equal contact angles (90°). For binary systems forming a shell of ~1 layer of melanin or silica on the supraball surface (not the core-shell morphology that is described at the end), we incorporated differing contact angles (100° for shell forming type and 80° for non-shell forming type) for melanin and silica to enable generation of the shell.[39,48] For systems with a repulsive interface, we applied a repulsive harmonic potential[70] with strength $\varepsilon_{repulsive\ wall}$ = 1.0 k$_B$T and a cutoff distance set to the nanoparticles' radius.

*4.2.2 Simulation Method*

For each system, the nanoparticles were randomly inserted into a spherical region with radius of ~13 μm generating an initial occupied volume fraction, $\varphi$, of 0.03. Langevin dynamics, implemented in the LAMMPS software package,[71] was employed to maintain system temperature and mimic solvent effects on the nanoparticle motion. The LAMMPS thermostat damping coefficient for all nanoparticles was scaled based on their relative mass and size to the smallest melanin nanoparticle in the system to ensure a similar implicit solvent viscosity for all nanoparticles.[72] A series of equilibration stages, as previously described,[39,48] were applied to ensure a fully equilibrated system before the system began shrinking. To model a shrinking spherical confinement, the confinement radius decreased linearly over the course of the simulation at a rate resulting in a desired Peclet number of ~1.0.[48] A simulation timestep of 0.0025 $\tau$ (where $\tau$ was the simulation time) was used during assembly until the close-packed supraball was formed near $\varphi \approx 0.6$.

The core-shell morphologies were generated starting from an assembled monodisperse supraball. We selected the core nanoparticles as the nanoparticles within a radial distance from the supraball center with the distance set to identify 50% of the nanoparticles to maintain the equal nanoparticle volume fraction. The shell nanoparticles were the remaining nanoparticles not identified as being in the core. The hexagonally packed crystal (HCP) supraball was generated by placing 220 nm diameter spheres in an HCP crystal and removing spheres that were beyond the desired supraball radius of ~5 μm. Supraballs with $\varphi < 0.6$ were generated by randomly removing nanoparticles from a close-packed supraball with $\varphi \approx 0.6$ until the desired $\varphi$ was achieved. All visualizations of the supraballs were performed using Visual Molecular Dynamics (VMD) software.[73]

**4.3 FDTD Simulations**

The optical reflectance of different types of supraball morphologies were simulated by performing three-dimensional FDTD calculations using a commercial-grade Ansys Lumerical 2021 R1 FDTD solver (Ansys, Inc). The FDTD method provides a general solution to any light scattering problem on arbitrary geometries by numerically solving Maxwell's curl equations (first principles) as derivatives with finite differences on a discrete spatial and temporal grid with user-defined resolution. By using a leap-frog approach, the electromagnetic waves (electric and magnetic field components) evolve iteratively through time. A typical FDTD simulation set-up as visualized in a CAD geometry can be viewed in **Figure S10**.

The CG-MD-simulated structures were imported into the solver and assigned corresponding material properties. The values of complex RI for silica and synthetic melanin were adapted from previous literature.[27,74] In order to replicate the conditions during experimental reflectance measurements using a microspectrophotometer that inscribes ~3 μm x 3 μm sensor area to collect reflected light at the center of the supraball, we decided to define the FDTD simulation region of the same lateral dimensions centered along the simulated supraball morphologies. The simulations were running at normal incidence using a broadband plane wave source (400 nm - 700 nm), propagating along the -Z direction. Boundary conditions in the lateral dimension (X and Y) were set to periodic. The reflectance data was collected using a Discretized Fourier Transform (DFT) power monitor set behind the source injection plane. We ensured that adequate simulation time (in fs) and boundary conditions along the light propagation direction (Z; perfectly matching layer (PML) boundaries) were chosen such that the electric field decayed before the end of the simulation (auto-shutoff criteria) and that all the incident light was either reflected, transmitted, or absorbed. A careful stepwise convergence testing (on parameters like proximity of PML boundaries, reflection from the PML, mesh sizes and accuracy, and source and monitor placements) was also performed to determine the variation in the reflectance spectra calculated from the numerical simulations. The variation ranged on the order of ~2-4% which was significantly less than the standard deviation observed from the experimental reflectance measurements (~10-15%).

In this study we used the following parametric values to set-up our optical simulations: a) an auto non-uniform mesh type with a mesh accuracy of 4 (18 mesh points per wavelength), minimum mesh step of 0.25 nm, and inner mesh size of ~12 nm for the structural part of the simulation box, b) a source injection plane at ~1.75 μm above the surface of the supraball, c) a stretched-coordinate PML boundary (steep-angle type) with 64 layers in the direction of propagation of incident light (Z plane) arranged ~1.25 μm behind the source injection plane, and d) a reflectance DFT monitor is set at ~0.75 μm behind the source injection plane. For all the simulations, the simulation time was set at 7500 fs and the auto-shutoff level (a rough estimate of the energy remaining in the simulation box as a fraction of power injected) was maintained at $10^{-5}$ to trigger the end of simulation upon achieving full decay. The simulated reflectance spectra presented throughout the study were obtained by averaging the results of optical modeling of three structures for each morphological type, simulated using both *p*- and *s*-polarization states of incident light with the error bars representing the standard deviations.

**4.4 Analyses**

Supraball structures were analyzed using the radial distribution function (RDF) and structure factor, S(q). The S(q) was calculated using the Debye equation.[75,76]

$$S_{\alpha\beta}(q) = \delta_{\alpha\beta} + \frac{1}{N_{\alpha\beta}} \sum_{i=1}^{N_\alpha} \sum_{j>i}^{N_\beta} \frac{\sin(qr_{ij})}{qr_{ij}} \qquad (1)$$

The $r_{ij}$ term was the radial distance between a pair of nanoparticles, $N_{\alpha\beta} = N_\alpha$ if α = β, and $N_{\alpha\beta} = N_\alpha + N_\beta$ if α ≠ β. The S(q) was calculated for the $S_{Mel-Mel}(q)$, $S_{Sil-Sil}(q)$, and $S_{Mel-Sil}(q)$ using only Melanin-Melanin pairs, Silica-Silica pairs, and Melanin-Silica pairs by adjusting the $N_\alpha$ and $N_\beta$ nanoparticles compared. The RDF and S(q) calculations were performed for the entire supraball and the portion of the supraball considered in the FDTD simulations to identify any structural differences between the analyses. **Figure S11** illustrates that there is a negligible difference between the S(q) and RDF from the entire supraball and the S(q) and RDF from the portion used in the FDTD simulations. The S(q)s were further analyzed to obtain primary S(q) peak heights, peak positions ($q_{pk}$), and peak widths (FWHM, $\Delta q$) by fitting the primary S(q) peak with Lorentzian function.

The simulated reflectance spectra acquired from FDTD calculations were analyzed to obtain key reflectance parameters like primary reflectance peak wavelengths, primary reflectance peak widths (FWHM), chromaticity coordinates, saturation, and lightness factors. To calculate primary reflectance peak wavelengths and peak widths, the simulated reflectance curves were first smoothed using local regression called locally weighted scatterplot smoothing (LOWESS). Igor Pro 8.04 (Wavemetrics, Inc.) was used to perform the smoothing and curve fitting routines. For smoothing, we chose the LOWESS algorithm with the regression order set to 1 (the routine fits for a line to the locally-weighted neighbors around each point) and value for the smoothing window was set to 0.1. For fitting the smoothed reflectance spectra, the Multipeak Fitting 2 package was used where the curves were fitted using a baseline of y = 0 and multi-Gaussian functions. The functional peak which provided the major contribution to the overall reflectance spectrum (in terms of reflectance intensity) was used to extract values of peak wavelengths and peak widths. For the calculating chromaticity coordinates, saturation, and lightness factors, the unsmoothed reflectance data were analyzed following the CIE 1931 and 1976 standards,[77] details of which can be found in the SI.

The statistical analyses on the datasets presented in this study were performed in JMP® Pro 16.0.0 (SAS Institute, Inc.). The datasets that were tested to evaluate if there was a relationship between independent and dependent variables underwent linear regression. We ensured that the basic assumptions of linear regression - equality of variances (Levene's test; $p > 0.05$) and normality of residual distribution (Shapiro-Wilk's test; $p > 0.05$) were met. For comparing between means of different treatment groups, one-way analysis of variance (ANOVA) was employed. If the test of equal variances was satisfied (Levene's test; $p > 0.05$), a parametric *post-hoc* test called Tukey HSD was performed to get multiple pair-wise comparisons between the means of the groups. When comparing only two groups, a standard Student's t-test was executed to test significant differences between the means. If the test of equal variances was not satisfied (Levene's test; $p < 0.05$), a non-parametric *post-hoc* test called Games-Howell multiple pairwise comparison test was performed. We used the non-parametric *post-hoc* test only for a few cases like one-way analysis of primary reflectance peak wavelength by packing fraction (**Figure S4a**), saturation by degree of mixing (**Figure 5e**), and primary reflectance peak wavelength and saturation by degree of stratification (**Figure 6d**; **Figure 6e**). The residuals of all the datasets were normally distributed (Shapiro-Wilk's; $p > 0.05$).

## 5. Supporting Information

- Description of color-defining parameters using CIE standards; Particle size and dispersity determination from neutron scattering; Supraball coloration and reflectance as melanin's absorption is adjusted for monodisperse and 20% dispersity; Packing fraction and nanoparticle dispersity impact on peak reflectance, saturation, and lightness factor; Comparison of attractive and non-attractive interface during supraball assembly; Structural information, S(q) and RDF, for all degrees of particle mixing; Polydispersity effect on the degree of mixing (strongly demixed and strongly mixed); Radial composition of silica nanoparticles for all type-based stratified supraballs; FDTD setup visualization; Structural difference, S(q) and RDF, between entire supraball and portion considered for FDTD simulations; CIE standards used in the calculation of tristimulus values.

## 6. Author Contributions

[‡]A.P. and C.M.H. contributed equally to this work. A.P., C.M.H., A.J., and A.D. contributed to project conceptualization and design of experiments. A.P. performed optical modeling (FDTD calculations) and C.M.H. conducted CG-MD simulations to generate supraball structures. Z.H. and N.C.G. performed synthesis of melanin and silica particles. A.P. fabricated one-component melanin and silica supraballs. B.V. and M.D.S. assisted in the reflectance measurements of one-component melanin and silica supraballs. A.P. and M.B. performed neutron scattering experiments at the SANS beamline, processed raw data, and completed data fitting. S.S. assisted in the data collection and processing during neutron scattering experiments. S.K.S. provided guidance during neutron scattering experiments, augmented conceptual understanding, and assisted in data processing and fitting. A.P. conducted all reflectance data processing and analyses, structural data analyses, color analyses, and statistical analyses. C.M.H. calculated all computational structure factors and RDFs. A.P. and C.M.H. wrote the initial manuscript. All authors reviewed and edited the manuscript. All authors have given approval to the final version of this manuscript.

## 7. Acknowledgment

A.P., C.M.H., S.S., Z.H., N.C.G., A.J., and A.D. acknowledge financial support from the Air Force Office of Scientific Research (AFOSR) under Multidisciplinary University Research Initiative (MURI) grant (FA 9550-18-1-0142). S.K.S acknowledges support from the Office of Basic Energy Sciences, U.S. Department of Energy under Department of Energy grant No. DE-SC0018086. B.V. and M.D.S. acknowledges support from AFOSR grant (FA9550-18-1-0477), FWO grant (G007117N), and HFSP grant (RGP 0047). This work was supported with computational resources from the University of Delaware (Farber and Caviness clusters) and the University of Akron (high performance computing workstations). A.P., S.S., M.B., S.K.S., and A.D. gratefully acknowledge the efforts of Cedric Gagnon and Jeff Kryzwon during data collection on the beamlines at NIST. Access to vSANS and NG7 SANS was provided by the Center for High Resolution Neutron Scattering (CHRNS), a partnership between the NIST and the National Science Foundation (NSF) under Agreement No. DMR-2010792. Specific commercial equipment, instruments, or materials are identified in this paper to foster understanding. Such identification does not imply recommendation or endorsement by the NIST, nor does it imply that the materials or equipment identified are necessarily the best available for the purpose. Finally, we extend our gratitude to the entire MURI Melanin team for helpful discussions and insights throughout the course of this project.

## 8. Notes

The authors declare no competing financial interest.